\documentclass[sigconf, nonacm]{acmart}

\AtBeginDocument{%
  \providecommand\BibTeX{{%
    Bib\TeX}}}
    \usepackage{comment}
\usepackage[inline]{enumitem}
\usepackage{amsmath,array,graphicx}
\usepackage{tcolorbox}
\usepackage{multirow}
\usepackage{tabularx}

\usepackage{pifont}
\usepackage{float}
\usepackage{xspace}

\usepackage{hyperref}

\AtBeginDocument{%
  \providecommand\BibTeX{{%
    \normalfont B\kern-0.5em{\scshape i\kern-0.25em b}\kern-0.8em\TeX}}}

\acmConference[ISSTA 2024]{ACM SIGSOFT International Symposium on Software Testing and Analysis}{16-20 September, 2024}{Vienna, Austria}

\newcommand{\cmark}{\ding{51}}%
\newcommand{\xmark}{\ding{55}}%
\newcommand{\RQOne}{\emph{What is the current state of reproducibility of scientific research in the domain of deep learning-based software fault prediction?}}

\newcommand{\RepositoryLink}{\textit{\textbf{Repository Link}}\xspace}
\newcommand{\ModelCode}{\textit{\textbf{Model Code}}\xspace}
\newcommand{\BaselineCode}{\textit{\textbf{Baseline Code}}\xspace}
\newcommand{\FinetuningCode}{\textit{\textbf{Hyperparameter-tuning Code}}\xspace}
\newcommand{\PreprocessingCode}{\textit{\textbf{Preprocessing Code}}\xspace}
\newcommand{\SupplementaryInfo}{\textit{\textbf{Supplementary Info.}}\xspace}

\newcommand{\ModelHyperparam}{\textit{\textbf{Model Hyperparam.}}\xspace}
\newcommand{\BaselineHyperparam}{\textit{\textbf{Baseline Hyperparam.}}\xspace}
\newcommand{\OptimizationProcedure}{\textit{\textbf{Optimization Procedure}}\xspace}
\newcommand{\ModelHyperparamRanges}{\textit{\textbf{Model Hyperparam. Ranges}}\xspace}
\newcommand{\BaselineHyperparamRanges}{\textit{\textbf{Baseline Hyperparam. Ranges}}\xspace}
\newcommand{\ModelBestHyperparam}{\textit{\textbf{Model Best Hyperparam.}}\xspace}
\newcommand{\BaselineBestHyperparam}{\textit{\textbf{Baseline Best Hyperparam.}}\xspace}

\newcommand{\PublicDataset}{\textit{\textbf{Public Dataset}}\xspace}
\newcommand{\DownloadableDataset}{\textit{\textbf{Downloadable Dataset}}\xspace}
\newcommand{\PreprocessedDataset}{\textit{\textbf{Preprocessed Dataset}}\xspace}
\newcommand{\PreprocessingSteps}{\textit{\textbf{Preprocessing Steps}}\xspace}
\newcommand{\MetadataDataset}{\textit{\textbf{Meta-data Info. Dataset}}\xspace}

\newcommand{\EvaluationProcedure}{\textit{\textbf{Evaluation Procedure}}\xspace}
\newcommand{\EvaluationMetrics}{\textit{\textbf{Evaluation Metrics}}\xspace}
\newcommand{\SignificanceTesting}{\textit{\textbf{Significance Testing}}\xspace}

\newcommand{\change}[2]{\textcolor{blue}{#2}}

\begin{document}

\title{Investigating Reproducibility in Deep Learning-Based Software Fault Prediction}

\author{Adil Mukhtar}
\affiliation{
  \institution{Graz University of Technology}
  \city{Graz}
  \country{Austria}
}\email{amukhtar@ist.tugraz.at}

\author{Dietmar Jannach}
\affiliation{%
  \institution{University of Klagenfurt}
  \city{Klagenfurt}
  \country{Austria}}
\email{dietmar.jannach@aau.at}

\author{Franz Wotawa}
\affiliation{%
  \institution{Graz University of Technology}
  \city{Graz}
  \country{Austria}
}\email{wotawa@ist.tugraz.at}
\renewcommand{\shortauthors}{Mukhtar et al.}

\begin{abstract}
Over the past few years, deep learning methods have been applied for a wide range of Software Engineering (SE) tasks, including in particular for the important task of automatically predicting and localizing faults in software. With the rapid adoption of increasingly complex machine learning models, it however becomes more and more difficult for scholars to reproduce the results that are reported in the literature. This is in particular the case when the applied deep learning models and the evaluation methodology are not properly documented and when code and data are not shared. Given some recent---and very worrying---findings regarding reproducibility and progress in other areas of applied machine learning, the goal of this work is to analyze to what extent the field of software engineering, in particular in the area of software fault prediction, is plagued by similar problems. We have therefore conducted a systematic review of the current literature and examined the level of reproducibility of 56 research articles that were published between 2019 and 2022 in top-tier software engineering conferences. Our analysis revealed that scholars are apparently largely aware of the reproducibility problem, and about two thirds of the papers provide code for their proposed deep learning models. However, it turned out that in the vast majority of cases, crucial elements for reproducibility are missing, such as the code of the compared baselines, code for data pre-processing or code for hyperparameter tuning. In these cases, it therefore remains challenging to exactly reproduce the results in the current research literature. Overall, our meta-analysis therefore calls for improved research practices to ensure the reproducibility of machine-learning based research.
\end{abstract}



\settopmatter{printfolios=true}
\maketitle

\section{Introduction}
\label{sec:introduction}
Deep learning techniques have become prevalent in Software Engineering (SE) related tasks~\cite{survey-se-dl} and applied in tools related to code summarization~\cite{iyer-etal-2016-summarizing, li2023classsum}, fault prediction~\cite{pornprasit2022deeplinedp, majd2020sldeep, hoang2019deepjit}, vulnerability prediction~\cite{10.1145/3468264.3468597, 10.1145/3540250.3558936} and more. This is mainly because of the adaptation and popularity of deep learning in other domains such as computer vision~\cite{voulodimos2018deep, o2020deep, wu2017application}, natural language processing~\cite{otter2020survey, young2018recent}, and more, have attracted researchers from the software engineering domain. Consequently, a large fraction of today's research is based on these learning approaches. Already back in 2018, Li et al.~\cite{li2018deep} reviewed 98 scientific articles from the software engineering domain and found that 84.7\% of the examined articles utilized some form of deep learning approaches for a multitude of software engineering tasks. While deep learning methods are popular and effective, their complex functioning requires precise documentation and detailed description of guidelines~\cite{abbess2022checklist, hartley2020dtoolai} to reproduce the results~\cite{haibe2020transparency}. Additionally, the availability of relevant code packages, including hyperparameter, evaluation, dataset, etc., is also a prerequisite for reproducibility.

Reproducibility has indeed become an important consideration in general AI research, and it is one of the main growing concerns in the scientific community~\cite{gundersen2018state, 10.1145/3477535}. Generally, reproducibility is about validating claims that are made in scientific papers and thereby ensuring progress in producing the same results claimed by the authors based on the provided documentation and information~\cite{gundersen2018state}. Researchers are often encouraged to share their artifacts, including datasets, executable files, and evaluation scripts, when publishing their findings. However, recent studies~\cite{Ghanta2018ASP,9041744, 10.5555/3454287.3454779} concerning deep learning indicate that a reproducibility crisis exists that undermines the reproducibility of the proposed work. Another assessment of the reproducibility status of deep learning models in the software engineering domain revealed that over 62\% of the published articles failed to provide quality code or comprehensive data, thus affecting the reproducibility of the presented research~\cite{10.1145/3477535}. Furthermore, Collberg and Proebsting inspected 601 source code repositories that were shared alongside scientific articles~\cite{10.1145/2812803}. The results revealed a rather modest success rate, with only 32.1\% of the experiments proving to be executable with success.

In general, reproducibility is considered a desirable characteristic. In the context of software engineering and artificial intelligence, it is expected that researchers provide the required documentation and code packages to reproduce the results because this helps other researchers validate the experimental results and findings. However, much of the scientific work fails in this regard. This situation extends beyond computer science and artificial intelligence research, as highlighted by researchers in the field of psychology, who have reported a success rate of just 36\% in reproducing the results reported in articles~\cite{doi:10.1126/science.aab2374}.
\begin{figure}[h!t]
    \centering
    \includegraphics[clip, trim=0cm 7cm 12.5cm 6.5cm, width=1.00\textwidth]{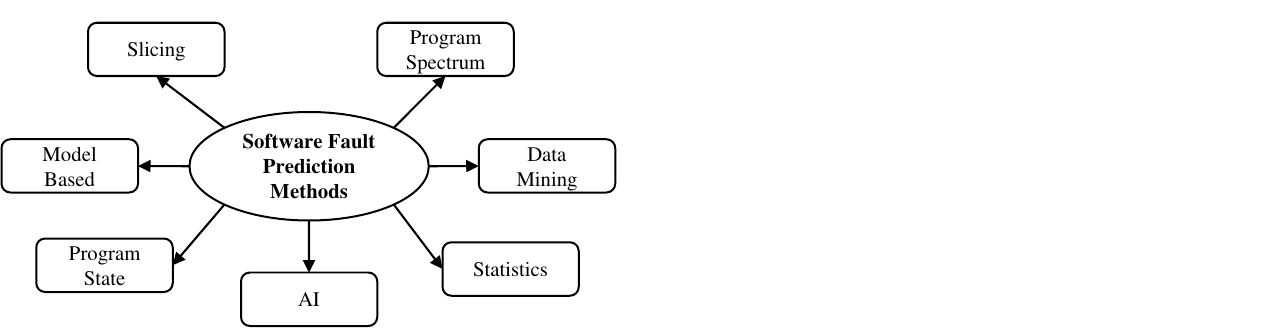}
    \caption{Categorization of Software Fault Prediction Techniques, adapted from~\cite{WongGLAW2016} }.
    \label{fig:fl-taxonomy}
\end{figure}
Therefore, our main objective is to investigate the current state of reproducibility in the domain of deep learning-based software engineering. We narrow our focus on fault prediction\footnote{In this survey, we treat bug prediction and defect prediction as forms of fault prediction.} techniques. The motivation for this investigation is based on several reasons. First, fault prediction is an essential and crucial task in software engineering, and numerous deep learning-based methods are proposed~\cite{qiao2020deep,mukhtar2023boostingspectrum,10.1145/3468264.3468597,10.1145/3540250.3558936,8668043}. Second, while existing studies on reproducibility analysis have considered tools written in a specific programming language, i.e., C++/C~\cite{Nong2023OpenSI} or have addressed software engineering tasks~\cite{10.1145/3477535} and artificial intelligence in general~\cite{gundersen2018state}, our research focuses on the fault prediction task of software engineering. Lastly, reproducibility in software engineering is often neglected or poorly documented~\cite{liu2018deep}. With this research, we hope to contribute to the current findings on the reproducibility status of deep learning-based software engineering. Hence, we formulate our research question as follows:

\begin{tcolorbox}
\RQOne
\end{tcolorbox}

To answer our research question, we conduct a systematic review in which we follow the definition of reproducibility provided by the Association for Computing Machinery (ACM)\footnote{\url{https://www.acm.org/publications/policies/artifact-review-and-badging-current}} and Gundersen~\cite{gundersen2018state}, i.e., reproducibility means that an independent group of researchers can obtain the same results claimed by the authors using the same methods, i.e., deep learning in our case, and measurement procedures provided by the authors. We selected research articles related to fault prediction in software engineering from top conferences published between 2019 and 2022. Then, we reviewed selected articles to record various reproducibility aspects proposed in the literature~\cite{gundersen2018state, abbess2022checklist}.

Our analysis revealed that approximately two-thirds of the research articles provide source code repositories. However, upon further inspection, we identified a significant gap in the availability of crucial details related to evaluation, model construction, hyperparameter optimization, and baseline methodologies. These critical aspects of information are absent in almost half of the articles, and in some instances, they are missing in more than half of the examined articles. Our investigation revealed that the scientific work related to deep learning-based SE for fault prediction tasks is not properly documented to a large extent to ensure reproducibility.

The paper is organized as follows: We discuss the background and related work in Section~\ref{sec:related-work} followed by our research methodology in Section~\ref{sec:reasearch-methodology}. Our results and findings are presented in Section~\ref{sec:results}. Finally, we discuss possible threats to validity and summarize  our research in Section~\ref{sec:threats-to-validity} and Section~\ref{sec:conclusion} respectively.

\section{Background \& Related Work}
\label{sec:related-work}
In this section, we provide a brief overview and discussion of the applicability of deep learning models in the context of software engineering. Subsequently, we present a summary of existing research that analyzes the current state of reproducibility.

\subsection{Deep Learning in Software Engineering}
\label{sec:background-related-work-softwaredebugging}
Over the years, the adoption of deep learning in software engineering has gained significant traction~\cite{10.1145/3477535, harman2012role, alshammari2022trends} and numerous deep learning methodologies have been introduced to address and improve various SE tasks, e.g., requirements analysis~\cite{navarro2017towards, madala2017automated}, clone detection~\cite{white2016deep, li2017cclearner, lei2022deep, fang2020functional}, debugging~\cite{huo2016learning, wong2011effective, qiao2020deep, wang2018software, mukhtar2023boostingspectrum}, and more.

In the context of requirements analysis, Navarro-Almanza et al.~\cite{navarro2017towards} introduced a Convolutional Neural Network (CNN) designed for the classification of functional and non-functional requirements. Similarly, Madala and colleagues introduced an automated approach for analyzing requirements, utilizing model-driven methods to autonomously identify component state transitions (CST)~\cite{madala2017automated}. Their approach incorporates Recurrent Neural Networks (RNN) with Long-Short Term Memory (LSTM) deep learning models. To assess the effectiveness of their model, they conducted a case study using a pacemaker requirements document.

Deep learning-based methods and tools for the software program clone detection task are also proposed. For instance, Li et al.~\cite{li2017cclearner} presented CCLEARNER to detect clones in a software program. CCLEARNER operates by leveraging method-level data associated with both clones and non-clones, extracting tokens from this information to train a deep-learning classifier. The tool's performance was assessed using a substantial benchmark of clones known as BigCloneBench. The approach underlying CCLEARNER involves feature extraction techniques, such as token categorization and similarity scoring among token vectors. The evaluation of the model and its comparison with existing clone detection methods revealed that CCLEARNER can efficiently detect clones while maintaining competitive performance levels. Likewise, Fang et al.~\cite{fang2020functional} proposed a novel technique that utilizes fusion embedding techniques to detect patterns and semantics associated with features of the source codes. Based on this information, a deep learning classifier is trained to detect clones in a large dataset of C++ programs. The outcomes of the experiments show that the proposed method significantly outperforms existing clone detection techniques.

Predicting and localizing faults are crucial aspects of software debugging, and numerous techniques are dedicated to fault prediction~\cite{WongGLAW2016}. Typically, these methods are classified into different categories depending on how they approach the task of fault prediction. This classification takes into account factors such as the involved program components and the characteristics of the used data, see Figure~\ref{fig:fl-taxonomy}. These techniques also often leverage deep learning models. Deep learning-based fault prediction techniques usually rely on the structural features of the underlying program and extract patterns that tend to correlate with the faultiness of the component. For instance, Huo et al.~\cite{huo2016learning} proposed a novel bug prediction technique named NP-CNN. This approach utilizes program structural information, bug reports, and source code to effectively identify buggy sections within the code. Qiao et al.~\cite{qiao2020deep} presented a fully connected neural network aimed at predicting the number of defects in program modules. The proposed technique leverages software metrics, such as the total number of code lines, the total number of characters, and Halstead's program length, etc. The assumption underlying this approach is that these metrics contain heuristic characteristics associated with the defective state.  Similarly, Mukhtar et al.~\cite{mukhtar2023boostingspectrum} presented a fault prediction technique for spreadsheet programs. They proposed to combine spectrum-based information with the characteristics of spreadsheet programs and utilize this information in a learning approach. Numerous other techniques have been proposed in the literature for fault prediction tasks~\cite{lv2016fault, manjula2019dnn, wang2020deepsemantic, zheng2016faultanalysis, batool2022software, alqasem2019dlfp, CHEN2020103, wong2011effective}.

\subsection{Reproducibility: A Growing Concern}
\label{sec:related-work-reproducibility}
In an extensively referenced \emph{Nature} news feature article titled \emph{``1,500 scientists lift the lid on reproducibility''}~\cite{Baker20161500SL}, Monay Baker presented findings from a survey revealing that more than 70\% of researchers reported that they attempted but could not reproduce the scientific work put forth by their peers. The survey showed that more than half of the researchers from various scientific disciplines such as chemistry, biology, medicine, computer science, and more, were unable to reproduce their own experimental results. This analysis provides an overview of the current status of reproducibility in scientific research as a whole, highlighting its emergence as a growing concern.

Gundersen et al.~\cite{gundersen2018state} targeted a specific domain, i.e., artificial intelligence, and reviewed 400 published articles from two highly-ranked conferences: International Joint Conference on Artificial Intelligence (IJCAI)~\footnote{\url{https://www.ijcai.org/}} and the Association for the Advancement
of Artificial Intelligence Conference (AAAI)~\footnote{\url{https://aaai.org/}}. The analysis revealed shortcomings in the adequate documentation of research published in these conferences. In particular, their findings indicated that nearly 70\% of the articles omitted clear articulation of research questions and the used methodologies. Additionally, a majority of the research papers exhibited non-reproducible outcomes. These were quantified through reproducibility variables covering aspects such as \emph{experiments}, \emph{data}, and \emph{method}. Remarkably, merely 25\% of these aspects were documented to facilitate reproducibility. In our present work, we follow the general direction of the investigation done by Gundersen et al.~\cite{gundersen2018state}, and we adapt various reproducibility variables used in their work. 

Within the software engineering domain, Nong and colleagues~\cite{Nong2023OpenSI} reviewed 55 deep learning-based vulnerability detection methods. Their investigation unveiled that only 25\% of the examined approaches offered publicly available tools. Furthermore, even among those works that did offer tools, there was a deficiency in documentation and implementation guidelines. This analysis aligns with the findings put forth by Gundersen et al.~\cite{gundersen2018state}, ultimately categorizing the state of the published research as largely non-reproducible. While vulnerability detection is a software engineering task that involves identifying vulnerable components within systems, we treat vulnerability detection as a form of fault prediction, as classified by Liu et al.~\cite{10.1145/3477535}, and we expand the scope to include articles concerning vulnerability detection within the context of fault prediction.

In the context of a comprehensive analysis of the reproducibility of deep learning methods in software engineering, Liu et al.~\cite{10.1145/3477535} surveyed 147 scientific articles presented at deep learning and software engineering conferences. The primary objective of this survey was to gain insights and assess the challenges associated with reproducibility.
The study findings revealed an important statistic: only 30\% of the examined articles published in software engineering conferences provided high-quality reproduction packages, whereas a notable 50\% of the articles published in AI venues provided such reproduction packages. They also found that only 10\% of the articles can address internal factors of models concerning reproducibility such as model stability, sensitivity to test data, and optimization convergence. In our review, however, we do not investigate these factors as we aim to record information such as links to code packages, hyperparameters, datasets, etc., provided by authors to ensure reproducibility. Another work on the reproducibility of defect prediction studies by Mahmood et al.~\cite{mahmood2018reproducibility} showed that only 6\% of the reviewed studies could be reproduced by other researchers.

Our investigation is partly similar to the state-of-the-art work presented by Gundersen et al.~\cite{gundersen2018state}. Our review is however more focused than previous works~\cite{mahmood2018reproducibility, gundersen2018state, Nong2023OpenSI, 10.1145/3477535}, targeting the particular task of fault prediction models. While our findings and implications may generally align well with such earlier works, our work provides an up-to-date perspective of the current status of reproducibility in an important subfield of software engineering. Importantly, our study aims to provide an in-depth characterization of what kind of documentation and software artifacts are nowadays provided by scholars to support the reproducibility of their work. Inherent factors of deep learning models such as randomness, test data sensitivity, and convergence may certainly affect reproducibility as well.  However, the analysis of such challenges is out of the scope of our present work, as we believe that explicit, clear, and correct documentation is a primary factor to be considered first when studying the potential of reproducibility of research work.

\section{Methodology}
\label{sec:reasearch-methodology}
We now discuss the methodology of our study. To begin, Section~\ref{sec:design} sketches the overall design of our study. Section~\ref{sec:selection-criteria} describes the selection process for scientific articles. Section~\ref{sec:data-extraction}, finally, explains how we measured the various aspects of reproducibility.

\subsection{Design}
\label{sec:design}

Our study is comprised of two main stages: \begin{enumerate*}
    \item article selection
    \item recording of reproducibility aspects
\end{enumerate*}. During the initial stage, we identify keywords related to fault prediction and deep learning to formulate search queries for the considered online research databases. We then apply our inclusion criteria to identify a subset of articles that is relevant to our research question. The detailed selection process is described in Section~\ref{sec:selection-criteria}.

In the second stage, we review the selected articles under various reproducibility aspects adapted from the existing literature and call these aspects \emph{reproducibility variables} as proposed by Gundersen et al.~\cite{gundersen2018state}. The variables are grouped into four categories: \emph{Source Code, Hyperparameter Tuning, Dataset, and Evaluation}. The main driving factor of the review process is to look for explicit documentation of information related to these variables in the articles and record measurements. The details of recording the measurements of these variables are further described in Section~\ref{sec:data-extraction}.

\subsection{Selection Criteria}
\label{sec:selection-criteria}
We apply a systematic retrieval process to ensure that the articles obtained from the online research database, namely \emph{IEEE Xplore}\footnote{\url{https://ieeexplore.ieee.org}} and \emph{ACM Digital Library}\footnote{\url{https://dl.acm.org/}}, are not only relevant but also within the designated date range, from 2019 to 2022. We selected these research databases because these are commonly used as a primary corpus in related studies, particularly in the field of deep learning-based software engineering applications~\cite{watson2022systematic, li2018deep, del2020trends}. Moreover, these databases provide access to millions of articles, such as journals, conferences, letters, magazines, and more~\cite{wilde2016ieee}.

\paragraph{\textbf{Relevant Terms}}
We consider terms relevant that are strongly related to the term ``fault prediction'', such as ``bug prediction'' and ``defect prediction''. We also consider the term ``fault localization'' relevant to our study since both fault prediction and fault localization techniques aim to locate faults~\cite{sohn2020bridging}. These terms are combined with the relevant deep-learning related keywords to expand our search retrieval. The keywords related to deep learning are presented in Table \ref{tab:dl-keywords}.

\begin{table}[!htpb]
\centering
\begin{tabular}{|c|c|}

\hline
deep learning  & artificial neural network \\ \hline
neural network & neural                    \\ \hline
decoder        & encoder                   \\ \hline
learning       & reinforcement learning    \\ \hline
reinforcement  & transformers              \\ \hline
\end{tabular}
\caption{Deep-learning-related keywords used in the study}
\label{tab:dl-keywords}
\end{table}
\begin{table*}[htbp]
\resizebox{\textwidth}{!}{%
\begin{tabular}{|c|c|c|}
\hline
\textbf{Conference Full Name}                                                                          & \textbf{Abbreviation} & \textbf{Ranking} \\ \hline
Automated Software Engineering                                                                    & ASE  & A*                 \\ \hline
European Software Engineering Conference and Symposium on the Foundations of Software Engineering & ESEC/FSE  & A*             \\ \hline
International Conference on Software Engineering                                                  & ICSE  & A*                \\ \hline
International Conference on Software Testing, Verification and Validation                         & ICST  & A                \\ \hline
International Conference on Software Analysis, Evolution and Reengineering                        & SANER & A                \\ \hline
International Conference on Software Maintenance and Evolution                                    & ICSME & A                \\ \hline
International Symposium on Software Testing and Analysis                                          & ISSTA   & A              \\ \hline
Mining Software Repositories                                                                      & MSR  & A                 \\ \hline
\end{tabular}
}
\caption{Top Software Engineering conferences selected for the study}
\label{tab:conferences}
\end{table*}
\paragraph{\textbf{Search String}}
After identifying keywords that correspond to the domain of deep learning-based software fault prediction methods, we formulate the search string as follows:
(``fault prediction'' OR ``bug prediction'' OR ``defect prediction'' OR ``fault localization'') AND (``deep learning'' OR ``neural network'' OR ``neural'' OR ``decoder''
                        OR ``encoder'' OR ``artificial neural network'' OR ``learning'' OR ``reinforcement'' OR ``reinforcement learning'' OR ``transformers'')

\paragraph{\textbf{Inclusion Criteria}} As our initial criterion, we exclusively consider articles published in SE conferences listed in Table~\ref{tab:conferences} as these are ranked top according to the CORE 2023 ranking\footnote{\url{https://www.core.edu.au/conference-portal}}. Moreover, these conferences are also often selected for survey studies~\cite{10.1145/3477535} in software engineering. The conference article is included if:
\begin{itemize}
    \item it uses deep learning for a fault prediction task,
    \item it is published between 2019 and 2022,
    \item it is published in a conference listed in Table \ref{tab:conferences}.
\end{itemize}

\begin{figure}[h!t]
    \centering

    \includegraphics[clip, trim=0cm 4.5cm 8cm 2.5cm, width=1.00\textwidth]{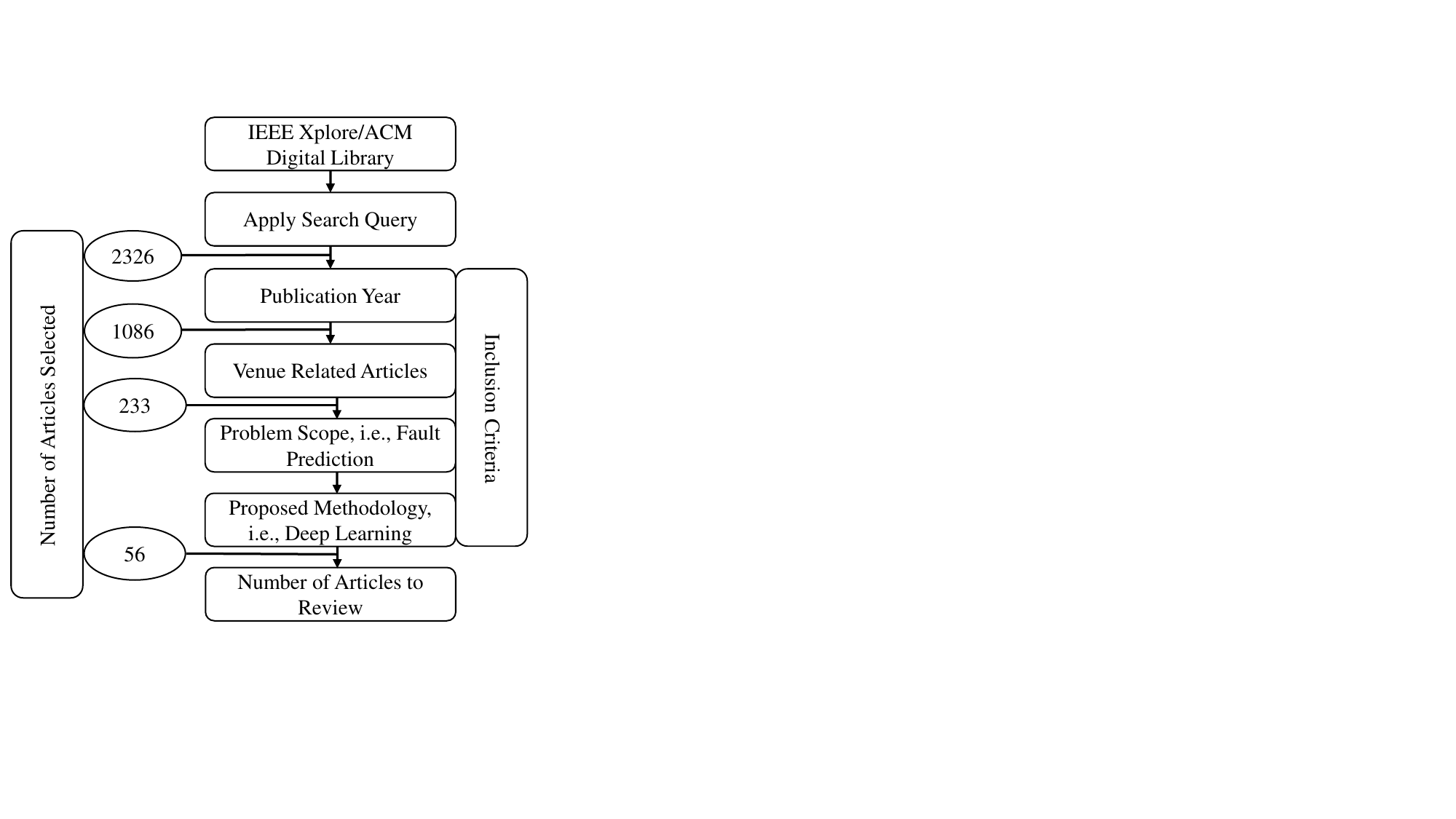}
    \caption{Articles selection process}.
    \label{fig:selection-process}
\end{figure}

As mentioned earlier, we also consider tasks related to \emph{vulnerability prediction} relevant to this study and consequently include those articles. As part of the exclusion criteria, we exclude articles not published in English and those that require payment to get access. However, in the end, we did not have to exclude any papers because the selected conferences only accept articles written in English, and because we had access to all proceedings through the university.
\paragraph{\textbf{Selection Process}}
Finally, we query the online research databases to retrieve articles and apply our inclusion and exclusion criteria. Articles that are discovered in more than one database are subjected to deduplication. The overall selection process of the articles is shown in Figure \ref{fig:selection-process}. After applying the search query, we initially identified 2,326 unique published articles associated with deep learning-based fault prediction tasks. Subsequently, by applying the publication year and venue-related criteria, we narrowed down the selection to 223 published articles. Finally, we conducted a manual review of each article to confirm that the article addressed a fault prediction task and utilized deep learning methods in its proposed methodology. This manual inspection resulted in 56 unique articles, which we then reviewed regarding reproducibility aspects. We share the list of selected articles along with the recorded reproducibility assessments online\footnote{{\url{https://bitly.ws/XNiz}}}.

\subsection{Reproducibility Variables}
\label{sec:data-extraction}
Table \ref{tab:reproducibility-vars} presents the reproducibility variables used in our study.
\begin{table*}[!htbp]
\resizebox{\textwidth}{!}{%
\begin{tabular}{cclc}
\hline
\textbf{}                                                                                  & \textbf{Variable}                    & \textbf{Description}                                                                                                                                         \\ \hline
\multirow{6}{*}{\textbf{Source Code}}                                                      & \RepositoryLink             & Is the link to an online source code repository provided and is the repository available?                                                                                                    \\ \cline{2-3} 
& \ModelCode                  & Is the source code for the proposed model(s) provided in the repository?                                                                                     \\ \cline{2-3} 
& \BaselineCode               & Is the source code for the baseline(s) provided in the repository?                                                                                          \\ \cline{2-3} 
& \FinetuningCode            & Is the source code for hyperparameter tuning provided in the repository?                                                                                     \\ \cline{2-3} 
& \PreprocessingCode          & Is the source code for data preprocessing provided in the repository?                                                                                        \\ \cline{2-3} 
& \SupplementaryInfo         & \begin{tabular}[c]{@{}l@{}}Is any supplementary information related to the source code provided,\\  e.g., README, Dockerfile, requirements.txt?\end{tabular} \\ \hline

\multirow{7}{*}{\textbf{\begin{tabular}[c]{@{}c@{}}Hyperparameter\\ Tuning\end{tabular}}} & \ModelHyperparam           & Is hyperparameter optimization discussed for the proposed model(s)?                                                                                     \\ \cline{2-3} 
& \BaselineHyperparam        & Is hyperparameter optimization discussed for the baseline(s)?                                                                                           \\ \cline{2-3} 
& \OptimizationProcedure      & Is the hyperparameter optimization procedure described?                                                                                                     \\ \cline{2-3} 
& \ModelHyperparamRanges    & Are the searched ranges for the hyperparameters reported for the proposed model?                                                                                          \\ \cline{2-3} 
& \BaselineHyperparamRanges & Are the searched ranges for the hyperparameters reported for the baselines?                                                                                               \\ \cline{2-3} 
& \ModelBestHyperparam     & Are the best hyperparameters reported for the proposed model?                                                                                                \\ \cline{2-3} 
& \BaselineBestHyperparam   & Are the best hyperparameters reported for the baselines?                                                                                                     \\ \hline
\multirow{5}{*}{\textbf{Dataset}}                                                          & \PublicDataset              & Is a pre-existing public dataset used?                                                                                                                                       \\ \cline{2-3} 
& \DownloadableDataset         & Are datasets available for download?                                                                                                                       \\ \cline{2-3} 
& \PreprocessedDataset        & Are preprocessed datasets available?                                                                                                                       \\ \cline{2-3} 
& \PreprocessingSteps        & Are the preprocessing steps described?                                                                                                                       \\ \cline{2-3} 
& \MetadataDataset     & Is meta-data information available for the dataset?                                                                                                      \\ \hline
\multirow{3}{*}{\textbf{Evaluation}}                                                       & \EvaluationProcedure        & Is the evaluation procedure described?                                                                                                                       \\ \cline{2-3} 
& \EvaluationMetrics                    & Are the metrics provided for the evaluation?                                                                                                                 \\ \cline{2-3} 
& \SignificanceTesting        & Are the details of statistical significance tests provided?                                                                                                              \\ \hline
\end{tabular}
}
\caption{Reproducibility variables and their descriptions}
\label{tab:reproducibility-vars}
\end{table*}
As part of the review process, we record these variables and look for clear mentions of information related to the variables. If the article provides a link to an online source code repository, we also thoroughly inspect the repository to verify whether further information related to the individual aspect can be found there.
During the review process, we use a checkmark (\cmark~) to indicate the presence of appropriate information regarding a specific variable, and we use the \xmark~ symbol to note its absence. We now explain the variables within each category.

\paragraph{\textbf{Source Code}}
Sharing of source code packages is considered an effective way to ensure reproducibility~\cite{gonzalez2012reproducibility, 10.1145/3477535}. In this category, we mainly look for the link to a source code repository. Ideally, the link should be provided within the article but in case it is not available, researchers could request the source code from the authors, e.g., through email. However, this is considered unreliable as suggested by Liu et al.~\cite{10.1145/3477535}, and we therefore only rely on the information and links provided in the articles. 
This inspection allows us to report the proportion of articles that share code packages. 

This category contains six variables as presented in Table \ref{tab:reproducibility-vars}.  During the review of the variable \RepositoryLink, we look for the link to an online source code repository and check if the link is available and the repository is available. We further inspect the repositories to locate the code scripts for the creation of proposed model(s) and baseline(s) when reviewing the variables \ModelCode and \BaselineCode, respectively. Similarly, we check if code is available for  dataset preprocessing and fine-tuning of the model while reviewing the variables \PreprocessingCode and \FinetuningCode. Additionally, we record a \cmark~if at least one of the files such as Dockerfile, README, and requirement.txt is found; otherwise, we record a \xmark~. 

\paragraph{\textbf{Hyperparameter Tuning}}
Deep learning models, like traditional machine learning models, usually have a number of hyperparameters that can be chosen depending on the given situation, such as embedding sizes, the learning rate, or weight factors for the loss function. The optimal values for such hyperparameters for a given dataset and optimization goal are commonly determined through a systematic procedure such as grid search. This tuning process is essential because the choice of the hyperparameters usually has a huge effect on the performance of a model \cite{shehzad2023everyone}. It is therefore highly important not only that all models in a comparison are properly tuned, but also that the tuning process is properly documented to ensure reproducibility, in particular in terms of searched parameter spaces, the search method, and the final parameters.

We identified seven variables that belong to the category of hyperparameter tuning-related information. For these variables, we primarily rely on the article content and pay particularly close attention to the sections where this information is expected to be usually available. First, we check if the discussion related to hyperparameter tuning is mentioned, i.e., fine-tuning for the proposed and baseline models and register values for the \ModelHyperparam and \BaselineHyperparam variables. Then we record whether the article describes the optimization procedure, such as random search or grid search, during the review of the \OptimizationProcedure variable. Similarly, we record if the information regarding parameter ranges and best parameters is presented for both the proposed and the baseline models. This is done while reviewing the \ModelHyperparamRanges, \BaselineHyperparamRanges, \ModelBestHyperparam, and \BaselineBestHyperparam variables. We also inspect the code repository in case the information is not found in the manuscript.

\paragraph{\textbf{Dataset}}
The availability of datasets is essential to reproduce reported results. However, datasets are often preprocessed before feeding into the deep learning models. For instance, log transformation and normalization are commonly applied as preprocessing methods. The preparation of the data is an important activity in developing effective deep-learning models. Therefore, the availability of such information is also equally important for reproducibility. Thus, dataset-related information is recorded through five reproducibility variables as presented in Table~\ref{tab:reproducibility-vars}. In many cases, methods are assessed using benchmark datasets, which are frequently publicly accessible. Sometimes, however, benchmark datasets are not used by authors and they create their own datasets. To record such information, we look for links to dataset sources and record the values for the \PublicDataset and \DownloadableDataset variables accordingly. The reason for including public datasets is that these datasets, e.g., PROMISE, NASA, are usually accessible over the web. However, these datasets are often available in some raw form and need to be processed according to the proposed model. Therefore, we check if the preprocessed dataset is accessible as well and record this in the \PreprocessedDataset variable. To record the information related to preprocessing steps, we inspect both the article content and supplementary information provided alongside the repository, and we correspondingly register a value for the \PreprocessingSteps variable. Moreover, we record a value for the \MetadataDataset variable to see if there is meta-information available pertaining to the dataset, including feature descriptions, feature ranges, and meta-level information. At times, authors reference previously published articles that contain datasets used in their own work. We also record this information, however, we only limit our investigation to the original article.
\begin{figure*}[h]
    \centering
    \includegraphics[clip, trim=5cm 6.7cm 5cm 7.5cm, width=1.00\textwidth]{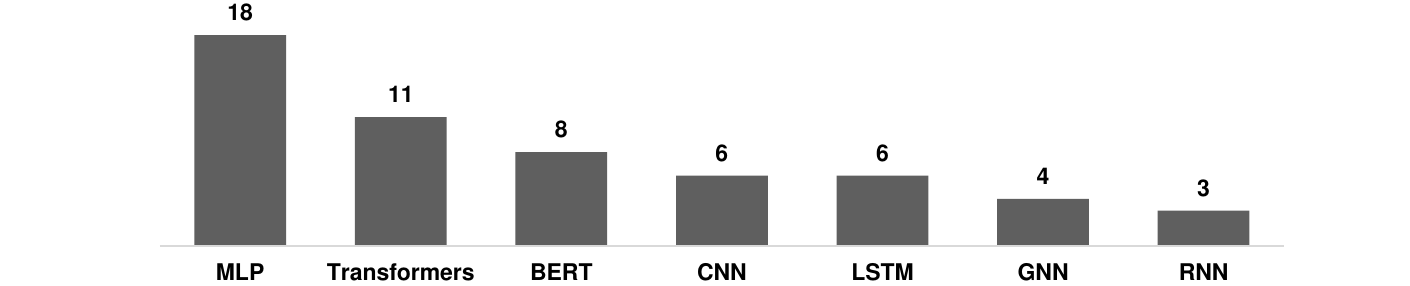}
    \caption{Frequencies of different deep learning architectures}
    \label{fig:dl-model-types-util}
\end{figure*}
\paragraph{\textbf{Evaluation}}
Deep learning methods are typically evaluated in an offline environment based on different evaluation protocols and based on various metrics. Commonly used evaluation protocols include k-fold cross-validation or leave-one-out evaluation; examples of evaluation metrics include precision and recall or the root mean squared error. An accurate description of these evaluation aspects is necessary for reproducibility. Therefore, we incorporate these aspects in our investigation and record whether the evaluation procedures and metrics are clearly explained. Additionally, we also take note if any statistical significance tests were reported in the article.

First, we record information concerning the evaluation methodology for the \EvaluationProcedure variable. Specifically, we look for the mentions and descriptions of the evaluation procedure. Then, we look for the evaluation metrics that are used to record a value for the \EvaluationMetrics variable. We also record a value for the \SignificanceTesting variable by checking if the article conducts significance testing to compare the performance of the proposed model with the baselines.

As an additional analysis, we also investigate the experimentation type used during the evaluation, i.e., offline or online. If the experimentation is done in an online setting then we record the following relevant information: \begin{enumerate*}
    \item Is the supplementary information provided related to user study, e.g., user manuals, documentation, etc.?
    \item What was the study size?
\end{enumerate*} 

Generally, please recall that throughout our investigation, our focus is to assess the availability of explicit documentation of information regarding reproducibility. Importantly, our goal is \emph{not} to retrain and evaluate the models using the information and materials provided by the authors. Instead, we exclusively focus on recording reproducibility variables (Table~\ref{tab:reproducibility-vars}) that indicate what the authors have provided in order to help reproducing the results they report in their papers.

We first present the results related to the analysis of the selected articles in Section \ref{sec:exploratory-analysis}. In Section \ref{sec:results}, we report and discuss the findings related to our main research question.

\subsection{Article Statistics}
\label{sec:exploratory-analysis}
Table \ref{tab:year-published} presents the yearly distribution of the selected articles. We can observe that the number of relevant articles increased sharply in 2022, while it remained rather constant in previous years. Generally, our sample may be too small---resulting from our specific focus---to derive reliable trends over time. Nonetheless, the uptake in 2022 stands out and indicates that deep learning models are becoming prevalent in top-tier software engineering conferences, and we found at least one relevant paper in each conference.

\begin{table}[H]
\centering
\caption{Number of articles published yearly}
\label{tab:year-published}
\begin{tabular}{c|c}
\hline
\textbf{Year Published} & \textbf{Number of Articles} \\ \hline
2019                    & 10                          \\
2020                    & 9                          \\
2021                    & 12                          \\
2022                    & 25
\\ \hline
\end{tabular}
\end{table}
\begin{figure*}[h!t]
    \centering
    \includegraphics[clip, trim=7cm 9.5cm 9cm 6.5cm]{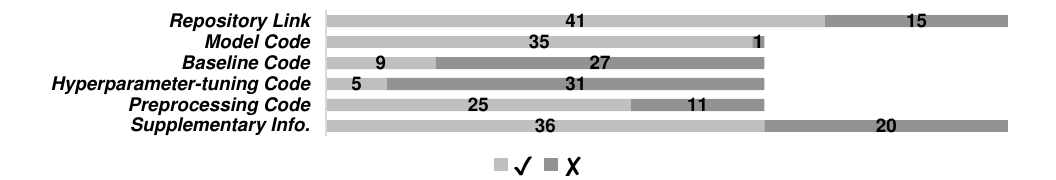}
    \caption{Analysis of Source Code Reproducibility Variables}
    \label{fig:source-code-analysis}
\end{figure*}
We furthermore recorded which deep learning architectures were used for fault prediction in the considered papers. Figure \ref{fig:dl-model-types-util} shows the distribution. Notably, the basic multilayer perceptron (MLP) network is the most frequently employed architecture. However, the much more recent \emph{transformer} architecture has been used very frequently, both in its standard form or as part of a specialized architecture like BERT (Bidirectional Encoder Representations from Transformers). The remaining articles show that researchers explored a rich range of popular architectures for the fault prediction task, including Convolutional Neural Networks (CNN), Long Short-Term Memory (LSTM), Graph Neural Networks (GNN), and Recurrent Neural Networks (RNN).

\section{Results and Findings}
\label{sec:results}

In this section, we provide the findings related to the reproducibility variables described in Table \ref{tab:reproducibility-vars}, considering both offline experiments and user studies.

\subsection{\textbf{Source Code}}
\label{sec:results-source-code-analysis}
The overall findings related to the source code variables are presented in Figure~\ref{fig:source-code-analysis}. We found that 41 articles, i.e., 73.2\%, share links to a source code repository. However, upon closer inspection, we noticed that 5 articles at the time of our analysis provided links to non-existent (potentially private) repositories or to empty repositories. Nonetheless, as there is at least some code for a bit less than two-thirds (64.3\%) of the cases, our findings suggest that researchers are aware of the importance of ensuring reproducibility of their work and put some effort into sharing their source code. 

The existence of the five non-working repository links may point to certain issues in the review and publication process. In these cases, reviewers were either not checking the repositories, or the authors only promised to share their code in case of acceptance, and did not share artifacts in the reviewing period. To address such issues, expectations regarding reproducibility aspects should be made explicit in the call for papers, and the existence of the shared artifacts must be verified before publication. 

Looking at the 36 articles with accessible repositories, we found that they contained the source code for the \emph{proposed} model (\textbf{\ModelCode} variable) in all but one case. In this particular case, we found that the code is provided for only one of the intermediate tasks, and not for the learning model. 

However, it turned out that the code for the models that were used as \emph{baselines} (\textbf{\BaselineCode} variable) in the empirical evaluations is missing in the large majority (75\%) of the cases. This is quite surprising because for a reliable evaluation, researchers must have access to the original source code of the baselines or have implemented them by themselves. In case the authors implemented the baselines by themselves, there is a certain risk that such implementations are not entirely correct and may lead to wrong conclusions, see \cite{DBLP:conf/recsys/HidasiC23}. Thus, it is important that the baseline code is shared as well. In case some original code is used, on the other hand, it must be ensured that both the baselines and the proposed model are benchmarked with some shared evaluation code. Otherwise, there is a risk that the comparison is not exact, because of certain differences, e.g., how datasets are split, how the metrics are computed, etc. In the literature, there are also some indications that sometimes authors do not actually run the baselines by themselves but simply copy the results from previous papers \cite{shehzad2023everyone}. Such an approach may however be very unreliable, because of the commonly observed differences in how the evaluation is done.
\begin{figure*}[ht]
    \centering
    \includegraphics[clip, trim=8cm 9.05cm 7cm 6.5cm]{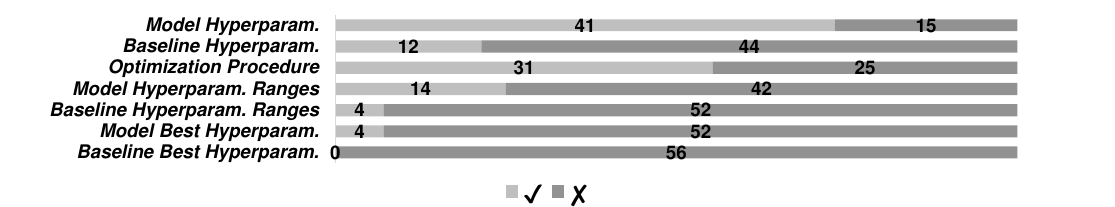}
    \caption{Analysis of Hyperparameter Tuning Reproducibility Variables}
    \label{fig:hyperparam-analysis}
\end{figure*}
Looking at the availability of code for the hyperparameter tuning process (\textbf{\FinetuningCode} variable), we find that this code is provided in only 13.8\% of the cases for which a repository with code was provided at all. Upon further inspection, we found a number of instances in which authors claimed to do hyperparameter tuning with a certain optimization method like Grid Search. However, the code for such procedures was then not included in the shared repository. Instead, we often found static hyperparameters in the shared model code, which could simply be the last ones that the authors tried before uploading their code.

In the context of hyperparameter tuning code, one might argue that this process can be accomplished with a few library calls\footnote{E.g., when using the scikit-learn Python library (\url{https://scikit-learn.org})}. However, the details of the process  (e.g., regarding the explored parameter ranges, the number of tries in a random search, etc.) can make a huge difference. Providing the tuning code can therefore be very important to be able to reproduce the exact results reported in a paper.

More positive observations were made regarding the availability of data preprocessing code (\textbf{\PreprocessingCode} variable), where more than two-thirds of the 36 accessible code repositories contained the respective code. Furthermore, all 36 repositories contained some at least minimal documentation (\textbf{\SupplementaryInfo} variable), e.g., in the form of a README file. However, only in four repositories we found a Docker\footnote{\url{https://www.docker.com}} file, which is assumed to be particularly helpful for easy reproducibility \cite{cremonesi2021aimag}, as such containers relieve researchers from the burden of setting up an execution environment with all needed software dependencies (in the right version).

\subsection{\textbf{Hyperparameter Tuning}}
\label{sec:results-hyperparameter-tuning}
Figure~\ref{fig:hyperparam-analysis} presents the statistics of the hyperparameter tuning reproducibility variables. We found that 41 (73\%) of the articles discuss hyperparameter optimization for the \emph{proposed} model (\textbf{\ModelHyperparam} variable). Among them, 31 articles explicitly provide the name of the optimization procedure (\textbf{\OptimizationProcedure} variable). When it comes to the optimization of the \emph{baseline} models (\textbf{\BaselineHyperparam} variable), in 78\% of the cases no information is provided, which aligns with the recent findings in \cite{shehzad2023everyone}, where researchers found that generally little documentation is provided about the tuning of baselines.

Looking at the documentation of explored hyperparameter ranges and the best model parameters (\textbf{\ModelHyperparamRanges} and \textbf{\ModelBestHyperparam} variables), we find this information is very frequently missing: 14 articles provide details for the hyperparameter ranges and only 4 articles report the best hyperparameter. Here, it is notable that \emph{none} of the articles report the best hyperparameter for the baselines, and a large number of articles, i.e., 94\%,  do not provide the searched ranges for the baselines. Again, this is surprising, because in case the authors had correctly tuned all baselines per dataset as required, questions regarding suitable ranges must have come up, and optimal hyperparameter values must have been determined at some stage.

Overall, we find that the \emph{documentation} of the details of the hyperparameter tuning process---in particular for the baselines---is largely missing in the analyzed literature. Clearly, we cannot know if the authors have actually properly tuned the baselines. In any case, tuning many baselines on several datasets can be computationally costly, in particular when it comes to modern deep learning architectures, and such a tuning process may take days or weeks. It is therefore a bit surprising that such efforts are not documented in a research paper at all. An alternative explanation therefore is that authors may sometimes spend extensive efforts to tune their proposed model, but do not apply the same efforts to the baselines. Such problematic research practices have been observed in other areas of applied machine learning earlier, e.g., in information retrieval, time series forecasting, or recommender systems \cite{Armstrong:2009:IDA:1645953.1646031,Lin:2019:NHC:3308774.3308781,Makridakis2018,ferraridacremaetal2019}. 
\begin{figure*}[h!t]
    \centering
    \includegraphics[clip, trim=8.3cm 9.1cm 8cm 7.2cm]{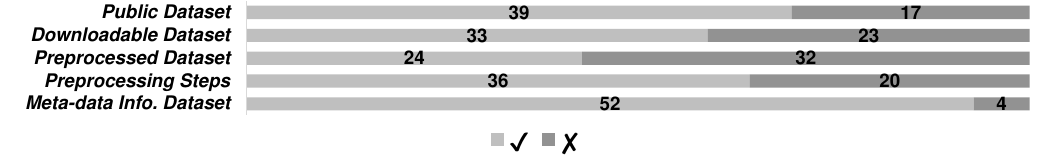}
    \caption{Analysis of Dataset Reproducibility Variables}
    \label{fig:dataset-analysis}
\end{figure*}

\subsection{\textbf{Dataset}}
\label{sec:results-dataset}
The findings for dataset-related reproducibility variables are presented in Figure~\ref{fig:dataset-analysis}. In a large majority of cases, around 70\%, researchers rely on pre-existing publicly available datasets in their experiments (\PublicDataset variable). The remaining 17 works used proprietary (new) datasets. In only one of these cases, the authors made their new dataset available by sharing a public link. Clearly, research works that are entirely based on non-public datasets cannot be validated or exactly reproduced by others.

In terms of the \emph{accessibility} of the used datasets (\textbf{\DownloadableDataset} variable), explicit links to the dataset files are provided in almost 60\% of the cases, either in the paper, the source code, or in some additional documentation.\footnote{In two cases, links were provided where either the link was broken or generically pointed to a conference website.} Dataset \emph{preprocessing} is very common in the reviewed papers. Almost two-thirds of the papers (36 out of 56) describe how the used datasets were preprocessed. The datasets that result from this preprocessing are however only shared in 24 of the papers. The reproducibility of the remaining 12 papers therefore depends on the ability of a researcher to exactly reproduce the preprocessing code. 

Finally, in almost all papers, at least some meta-information about the datasets was provided (\textbf{\MetadataDataset} variable), with the level of detail varying strongly across articles. Overall, however, it shows that the reproducibility of the reviewed works may frequently suffer from the use of non-public datasets and insufficiently documented data preprocessing procedures.

\change{\begin{table}[H]
\centering
\caption{Analysis of Dataset Reproducibility Variables}
\label{tab:dataset-analysis}
\begin{tabular}{c|c|c}
 & \cmark  & \xmark  \\ \hline
\textbf{\PublicDataset} & 39 & 17 \\ \hline
\textbf{\DownloadableDataset} & 33 & 23 \\ \hline
\textbf{\PreprocessedDataset} & 24 & 32 \\ \hline
\textbf{\PreprocessingSteps} & 36 & 20 \\ \hline
\textbf{\MetadataDataset} & 52  & 4 \\ \hline
\end{tabular}
\end{table}}{}

\subsection{\textbf{Evaluation}}
\label{results:evaluation}
We finally turn our attention to the documentation of evaluation aspects. Given their technical nature, all reviewed papers provide some empirical results and correspondingly describe their evaluation methodology. Notably, all except one article in our survey are exclusively based on \emph{computational} (offline) experiments; only one paper aimed to assess the proposed model's practical utility by involving developers in the fault prediction tasks. Unfortunately, in this single case the documentation of certain aspects of the study, e.g., participant demographics, was not very detailed.

\begin{figure}[htp]
    \centering
    \includegraphics[clip, trim=0cm 9cm 13cm 6.2cm, width=1.00\textwidth]{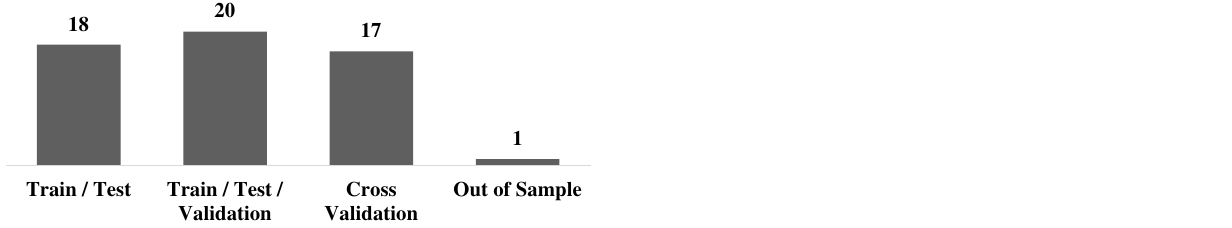}
    \caption{Evaluation methods used in the articles}
    \label{fig:analysis-evaluation-procedure}
\end{figure}

Different procedural approaches (\textbf{\EvaluationProcedure} variable) of different complexity are used in the papers for offline experiments, see Figure~\ref{fig:analysis-evaluation-procedure}. About one-third of the paper relies on single train-test splits, another third also includes a validation set for hyperparameter tuning, and about another third of the papers use a cross-validation procedure. We recall here that relying on one single train-test split may be considered risky or problematic for different reasons. First, it may imply that the hyperparameters were tuned on the test set---is not uncommon also in other areas of applied machine learning \cite{sun2020we}---but methodologically unsound. Second, a single (random) train-test split may have been created in an ``unlucky'' way so that the findings for this split do not generalize well. In this case, the code for creating the data splits must be shared by the authors to ensure reproducibility. Nevertheless, splitting techniques are the most common procedures for the evaluation. However, sometimes models are tested with data that is not part of the sample dataset, this is called out-of-sample evaluation and considered less sensitive to bias and variance~\cite{tantithamthavorn2016empirical}. Looking at Figure~\ref{fig:analysis-evaluation-procedure}, we see that it is applied in only one article as this is not a commonly used evaluation procedure.

A rich variety of metrics are used in the context of offline experiments, and all papers describe the metrics they use and the values obtained in the experiments (\textbf{\EvaluationMetrics} variable).  The most frequently used metrics are general ones that are commonly used for classification or ranking tasks such as precision, recall, area under the curve, and mean average precision. Additionally, for the fault prediction task, various metrics are used such as ``number of bugs detected'', relative improvement (\emph{RImp})~\cite{debroy2010grouping} or the EXAM score~\cite{mao2014slice} and other commonly used metrics. Finally, statistical significance tests are rather scarcely used in the reviewed papers (\textbf{\SignificanceTesting} variable), and only 8 papers provided outcomes of applying statistical significance tests.

\section{Implications}
\label{sec:discussions-implications}
Overall, while our study reveals some major open gaps in terms of reproducibility, we believe that the practices in the research community are developing in the right direction. Compared, for example, to the analysis by Gundersen et al.\cite{gundersen2018state} from 2018 in the broader field of AI, the fraction of papers for which code is shared is much higher in our survey than it was found by Gundersen et al. Also, looking even further back in history, sharing any code and data, to our experience, was not very common in many fields of computer science ten or fifteen years ago. As of today, many conferences and journals at least mention reproducibility aspects in their guidelines for authors and reviewers.   

Nonetheless, the situation is far from being ideal today, as our survey shows. For the majority of the published research, at least the code for the proposed models is shared. This certainly helps to \emph{reproduce the numerical results reported for the proposed models}, at least when other important information like datasets, preprocessing code, and optimal hyperparameters are available as well. A major problem here, however, is that in almost all cases the artifacts are missing to \emph{reproduce the claimed improvement over existing models}, as we, for example, cannot validate if the baselines were properly tuned. If this is not the case, we may observe a lot of ``phantom progress'' \cite{ferraridacremaetal2019} in the published literature and ``improvements that don't add up'' \cite{Armstrong:2009:IDA:1645953.1646031}.

To increase the level of reproducibility, we must analyze the potential reasons why scholars share or do not share artifacts, and we must subsequently develop suitable mechanisms that increase the probability of making research results more easily reproducible than today. Most intuitively, researchers might share their code because they are aware of the importance of reproducibility and they are seeking to report reliable and trustworthy results. Also, given that reproducibility aspects are increasingly considered explicitly in the academic reviewing process, sharing code may also increase their chances of paper acceptance. Finally, researchers may expect that their approaches are cited more often when others can easily use them, e.g., as baselines in their own experiments. 

On the other hand, creating all the materials that are needed for the reproducibility of the entire experiment can be very time-consuming.\footnote{In our own experience, software development for machine learning models and for research papers, in general, can be quite unstructured, leading to chaotic code organization or dead fragments because of time pressure or because of limited experience by the programmer.} Therefore, if the chances to get a research work published without sharing any materials or by sharing only one central component, it may be tempting for a researcher to avoid this extra work. Furthermore, some researchers may also fear that there are issues with their code artifacts, which may be exposed if they share their code. Further reasons for not sharing the code may exist as well.

We can think of different measures to further increase the level of reproducibility of published research in machine-learning-based software engineering. One measure is to make reproducibility more often one of the central criteria for the evaluation of research papers. This could be implemented through reproducibility guidelines published along with journal submission guidelines and conference calls for papers.\footnote{Various guidelines and checklists for machine learning and general AI exist, e.g., \cite{JMLR:v22:20-303} or \url{https://aaai.org/conference/aaai/aaai-23/reproducibility-checklist/}} Furthermore, reproducibility criteria should be made part of review forms so that the topic is explicitly checked and evaluated by reviewers. In terms of positive incentives for reproducible research, more and more computer science conferences nowadays have reproducibility tracks. Furthermore, organizations like ACM offer special recognition (badges) for reproducible research works\footnote{\url{https://www.acm.org/publications/artifacts}}.

Finally, and in our view most importantly, we believe that in the long run, existing issues of reproducibility and, consequently, of somewhat limited progress, can only be sustainably addressed through further increased awareness and education, involving all relevant stakeholders from students, teachers, scholars, or grant reviewers, see \cite{bauer2023overcoming}. Along with this, a shift in the academic incentivization and credit system seems required, one that values reproducibility much more than it is the case today.

\section{Threats to Validity}
\label{sec:threats-to-validity}

We recall that the articles we reviewed were retrieved from two specific research databases. One may thus argue that the selected sample of articles may not be fully representative. However, the rationale behind our choice of these particular databases is that these are commonly utilized in the academic literature for conducting surveys related to deep learning in software engineering~\cite{li2018deep, watson2022systematic, del2020trends}, and they offer access to a vast repository of articles \cite{wilde2016ieee}. With the goal to comprehensively evaluate the latest research works, we focused on articles published during the last four years. This selection criteria resulted in 56 articles that were considered in our review. While the number of research works published in the last decade is certainly higher, we believe that our review is well representative of the current state of research in terms of reproducibility.

We furthermore note that our research approach involves documenting exclusively what is reported and provided by the authors. We do not engage in the actual training and evaluation of the models. Furthermore, we do not take into account other factors affecting reproducibility, such as randomness or the non-deterministic characteristics of hardware~\cite{towards-training-rp-deepdl}. To efficiently review hundreds of articles, our effort is primarily toward the explicit documentation of reported information. To ensure the correctness of the recorded variables, we performed two rounds of article reviewing.

\section{Conclusion}
\label{sec:conclusion}
Deep learning techniques are gaining widespread popularity in the field of Software Engineering. Their constantly increasing complexity can however make it difficult to reproduce the results that are reported in research papers. To ensure progress in this field, it is therefore of utmost importance that scholars provide as much material as possible for others to validate and improve on their findings. Our survey revealed that the current situation in the subfield of software fault prediction is more than ideal. We as a community must constantly strive to further improve our standards in this important aspect of reproducibility, which is certainly a cornerstone of scientific research.

\section*{Acknowledgement}
This work is funded by the Austrian Science Fund
(FWF) under contract number P 32445.
\bibliographystyle{ACM-Reference-Format}
\bibliography{references}
\end{document}